\documentclass[aip,graphicx,secnumroman,eqsecnum]{revtex4-2}
\usepackage{graphicx}% Include figure files
\usepackage{dcolumn}% Align table columns on decimal point
\usepackage{bm}% bold math

\newcommand{\beq}{\begin{equation}}
\newcommand{\eeq}{\end{equation}}
\newcommand{\beqa}{\begin{eqnarray}}
\newcommand{\eeqa}{\end{eqnarray}}
\newcommand{\vc}[1]{\mbox{\boldmath $#1$}}

\newcommand{\vol}[1]{{\bf #1}}

\newcommand{\du}[1]{{\bf\sf #1}}

\begin{document}

%\preprint{APS/123-QED}

\title{Effect of retarded friction and added mass on the swimming speed of a vibrating two-sphere}

% Force line breaks with \\

%
\author{B. U. Felderhof}
 %\altaffiliation[Also at ]{Physics Department, XYZ University.}%Lines break automatically or can be forced with \\

\email{ufelder@physik.rwth-aachen.de}
\affiliation{Institut f\"ur Theorie der Statistischen Physik \\ RWTH Aachen University\\
Templergraben 55\\52056 Aachen\\ Germany\\}

\date{\today}% It is always \today, today,
             %  but any date may be explicitly specified

\begin{abstract}
A theoretical expression is derived for the mechanical contribution to the mean swimming speed of a vibrating two-sphere immersed in a viscous incompressible fluid. The two spheres are connected by an elastic spring which provides a harmonic potential for oscillations about a mean distance between centers. The system is made to oscillate at a chosen frequency by activating forces which sum to zero. The mechanical contribution to the resulting mean swimming velocity is calculated from the mechanical equations of motion and the corresponding impedance matrix of linear response. The frequency-dependent pair friction coefficients are found from approximate expressions derived earlier. The mechanical contribution is calculated to second order in the amplitude of stroke as a function of the scaling number, a dimensionless combination of size, frequency, and kinematic viscosity. Retarded friction and added mass determine the functional behavior.
\end{abstract}
%\pacs{47.15.G-, 47.63.mf, 47.63.Gd, 47.63.M-}% PACS, the Physics and Astronomy
                             % Classification Scheme.
%\keywords{Suggested keywords}%Use showkeys class option if keyword
                              %display desired
\maketitle
\section{\label{I}Introduction}
The swimming velocity of a deformable body in a fluid can be decomposed as a sum of two contributions.The first contribution follows from the equations of classical mechanics´for the individual body parts. These equations involve friction and added mass effects. The second contribution follows from the Navier-Stokes equations for fluid velocity and pressure. Both contributions are a result of rectification due to nonlinear interactions. We call the first contribution mechanical and the second one hydrodynamic. In the present article we study the first, mechanical contribution for a vibrating two-sphere immersed in a viscous incompressible fluid.

The work of Klotsa et al.\cite{1} established the vibrating two-sphere as  an important model in the theory of swimming. The model is surprisingly complex, but sufficiently simple to allow some mathematical analysis. Its partial solution can improve understanding and elucidate principles. This will be helpful in the analysis of more sophisticated swimmers.

In the case of the two-sphere the body parts are the two spheres.
The pair mobilities can be evaluated with various degrees of accuracy \cite{2}. Hubert et al. \cite{3} calculated the mean swimming speed of a two-sphere with Oseen type interactions for large distance between centers. For small amplitude of stroke we generalized this result to shorter distance \cite{4} and included added mass effects in dipole approximation. We used a kinematic approach in which the relative distance between centers is prescribed to oscillate harmonically. The mean speed was obtained as a function of the oscillation frequency.

In the following we study the two-sphere chain with harmonic elastic interaction, set into motion by periodic activating forces, which sum to zero. First we consider instantaneous hydrodynamic interactions without specifying the distance dependence and derive a novel expression for the mean speed of small amplitude swimming in terms of the impedance matrix of linear response theory. For Oseen type hydrodynamic interactions, with or without dipolar added mass effects, the expression for the mean speed reduces to that derived earlier by expansion in harmonics \cite{4}.

Next we generalize the theory to frequency-dependent friction coefficients, as first developed by Stokes for a single sphere on the basis of the linearized Navier-Stokes equations \cite{5}. It was realized by Basset \cite{6} that this corresponds to retarded friction, as described by a memory term in the equation of motion for the sphere \cite{6}$^,$\cite{7}. We use also retarded hydrodynamic pair interactions \cite{8} in the impedance matrix.

The second contribution to the swimming velocity results from flow generated by the swimmer via the movement of the spheres and via Reynolds stress. Since we need only the time-averaged swimming velocity, it suffices to solve the steady-state Stokes equations \cite{9}. The generated net flow is called the effect of steady streaming \cite{10}. We postpone the study of its contribution to the mean swimming veloity to a separate article.

Here we consider specifically two different systems, as described by four models of increasing sophistication. In the first system the two spheres are neutrally buoyant. This is indicated by $N$. In the second system the larger sphere is again neutrally buoyant, but the                  smaller sphere is loaded such that it has twice the mass of the larger sphere. This system is indicated by $L$. In the first model the hydrodynamic interactions between spheres are treated in Oseen-approximation. This is indicated by $O$, and is the model studied by Hubert et al. \cite{3}. In the second model we take approximate account of added mass effects by including the single sphere added mass in the equations of motion. The model is called the Oseen $^*$-model, and is indicated by $S$. The third model takes account of pair added mass effects in dipole approximation \cite{11}. This model is indicated by $D$. In the fourth model the friction coefficient of a single sphere is replaced by the frequency-dependent coefficient introduced by Stokes \cite{5}, and the pair hydrodynamic interactions are also made frequency-dependent, corresponding to retarded hydrodynamic interactions \cite{8}. In addition the single sphere friction coefficient is corrected for a pair effect such that the added masses are identical with those of the dipole model. This is called the retarded friction model, and is indicated by $R$. Thus $NR$ refers to the neutrally buoyant system described by the retarded friction model.

\section{\label{II}Mechanics with simple friction}

We consider two spheres of radii $a$ and $b$ and mass densities $\rho_a$ and $\rho_b$ immersed in a viscous incompressible fluid of shear viscosity $\eta$ and mass density $\rho$. The fluid is of infinite extent in all directions. We assume that at all times the centers are located on the $z$ axis of a fixed Cartesian system of coordinates. The dynamics of the system is governed by a harmonic interaction potential $V_{int}=\frac{1}{2}k(z_2-z_1-d)^2$, depending on the instantaneous configuration of centers $z_1(t),z_2(t)$, and by actuating forces $E_1(t)$ and $E_2(t)$ in the $z$ direction and varying harmonically in time with frequency $\omega$. We assume in this section that friction with the fluid can be described  by a $2\times 2$ friction matrix with elements which depend on distance $z(t)=z_2(t)-z_1(t)$, but are independent of frequency $\omega$. Added mass effects are described by a $2\times 2$ mass matrix $\du{m}(z)$, calculated from potential theory.

In the course of time the motion of the system becomes periodic with period $T=2\pi/\omega$ and with a steady mean swimming velocity $\overline{U}_{sw}$, the average being over a period. In the following we derive a simple expression for $\overline{U}_{sw}$ for the case of small amplitude harmonic variation of the activating forces $E_1(t)=\varphi_1\cos(\omega t)$ and $E_2(t)=-E_1(t)$. The mean velocity will be calculated to second order in the amplitude $\varphi_1$.

We summarize the positions of centers in the $2$-dimensional vector $\du{R}=(z_1,z_2)$, and the sphere momenta in $\du{p}=(p_1,p_2)$. The momenta are related to the velocities $\du{U}=(U_1,U_2)$ by $\du{p}=\du{m}\cdot\du{U}$ with a symmetric $2\times 2$ mass matrix $\du{m}$ which depends on distance $z$. The dynamics of the system is assumed to be governed by the approximate equations of motion \cite{12}
\begin{equation}
\label{2.1}\frac{d\du{R}}{dt}=\du{U},
\qquad\frac{d\du{p}}{dt}=-\frac{\partial\mathcal{K}}{\partial\du{R}}-\vc{\zeta}\cdot\du{U}-\frac{\partial V_{int}}{\partial\du{R}}+\du{E},
\end{equation}
where the kinetic energy $\mathcal{K}$ is given by
\begin{equation}
\label{2.2}\mathcal{K}=\frac{1}{2}\;\du{p}\cdot\du{w}\cdot\du{p},
\end{equation}
with inverse mass matrix $\du{w}=\du{m}^{-1}$. The friction matrix $\vc{\zeta}$ and the interaction potential $V_{int}$ are invariant under translations.

We abbreviated $\du{E}=(E_1,E_2)$. We assume $E_2(t)=-E_1(t)$, thus ensuring that the total applied force vanishes at any time. Another choice would imply that the center of mass of the swimmer is swung back and forth by an applied force during its motion. The activating forces cause internal motions of the body without acceleration- of its center of mass.

The total sphere momentum $P=p_1+p_2$ can be expressed as $P=\du{u}\cdot\du{p}$ with vector $\du{u}=(1,1)$. The momentum varies in time due to friction with the fluid as
 \begin{equation}
\label{2.3}\frac{dP}{dt}=-\du{u}\cdot\vc{\zeta}\cdot\du{U}.
\end{equation}
In the asymptotic periodic swimming situation we therefore have
\begin{eqnarray}
\label{2.4}\du{u}\cdot\overline{\vc{\zeta}\cdot\du{U}}=0,
\end{eqnarray}
where the overline bar indicates the average over a period. This shows that in periodic swimming the mean drag vanishes.

We perform a formal expansion in powers of the amplitude of the activating forces. To second order in the amplitude Eq. (2.4) reads
\begin{equation}
\label{2.5}\du{u}\cdot\overline{\vc{\zeta}^{(1)}\cdot\du{U}^{(1)}}+\du{u}\cdot\vc{\zeta}^0\cdot\overline{\du{U}^{(2)}}=0,
\end{equation}
where $\vc{\zeta}^0$ is the friction matrix of the rest configuration, which is time-independent.
The sphere velocities $\du{U}=(U_1,U_2)$ can be expressed as
\begin{equation}
\label{2.6}\du{U}=U\du{u}+\dot{\du{d}},
\end{equation}
where $U=(U_1+U_2)/2$ is the center velocity, and $\du{d}=(d_1,d_2)$ are the deviations from the co-moving rest positions. In the periodic swimming situation we have
\begin{equation}
\label{2.7}\overline{\du{U}}=\overline{U}\du{u}.
\end{equation}
From Eq. (2.5) we find for the mechanical mean second order swimming velocity
\begin{equation}
\label{2.8}\overline{U^{(2)}}_M=\frac{-1}{Z_0}\;\du{u}\cdot\overline{\vc{\zeta}^{(1)}\cdot\du{U}^{(1)}},
\end{equation}
where
\begin{equation}
\label{2.9}Z_0=\du{u}\cdot\vc{\zeta}^0\cdot\du{u}
\end{equation}
is the friction coefficient of the rest configuration.

The average on the right-hand side of Eq. (2.8) can be evaluated from linear response theory.
The first order friction matrix can be expressed as
\begin{equation}
\label{2.10}\vc{\zeta}^{(1)}=\du{d}^{(1)}\cdot\nabla\vc{\zeta}\big|_0,
\end{equation}
where $\nabla=(\partial/\partial z_1,\partial/\partial z_2)$. With activating forces $E_j(t)=\mathrm{Re}[E_{j\omega}\exp(-i\omega t)],\;(j=1,2)$ the displacements in linear response take the form
\begin{equation}
\label{2.7}d^{(j)}_j(t)=d_{jc}\cos(\omega t)+d_{js}\sin(\omega t),\qquad j=(1,2),
\end{equation}
where $d_{jc}=\mathrm{Re}[d_{j\omega}],\;d_{js}=\mathrm{Im}[d_{j\omega}]$.
The velocities are
\begin{equation}
\label{2.12}U^{(1)}_j(t)=U_{jc}\cos(\omega t)+U_{js}\sin(\omega t),\qquad j=(1,2).
\end{equation}
The bilinear averages over a period are
\begin{equation}
\label{2.13}\overline{d^{(1)}_jU^{(1)}_k}=\frac{1}{2}(d_{jc}U_{kc}+d_{js}U_{ks}).
\end{equation}
We evaluate these quantities from the linear response.

\section{\label{III}Correlations}

From the equations of motion Eq. (2.1) we find for the one-dimensional situation with coordinates $q_1=z_1,\;q_2=z_2-d$ and velocities $U_1,\;U_2$ after linearization
\begin{eqnarray}
\label{3.1}\frac{dq_1}{dt}&=&U_1,\qquad \frac{dq_2}{dt}=U_2,\nonumber\\
m_{11}\frac{dU_1}{dt}&+&m_{12}\frac{dU_2}{dt}+\zeta_{11}U_1+\zeta_{12}U_2+k(q_1-q_2)=E_1(t),\nonumber\\
m_{12}\frac{dU_1}{dt}&+&m_{22}\frac{dU_2}{dt}+\zeta_{12}U_1+\zeta_{22}U_2-k(q_1-q_2)=E_2(t),
\end{eqnarray}
with constant coefficients taken at the equilibrium distance $z=d$. For oscillating activating forces at frequency $\omega$ it is convenient to use complex notation. The equation for the amplitudes $\vc{\psi}_\omega=(q_{1\omega},q_{2\omega},U_{1\omega},U_{2\omega})$ then takes the form
\begin{equation}
\label{3.2}\vc{M}(\omega)\cdot\vc{\psi}_\omega=\vc{\varphi}_\omega,
\end{equation}
with matrix
\begin{equation}
\label{3.3}\vc{M}(\omega)=\left(\begin{array}{cccc}-i\omega&0&-1&0\\0&-i\omega&0&-1\\k&-k&-i\omega m_{11}+\zeta_{11}&-i\omega m_{12}+\zeta_{12}\\-k&k&-i\omega m_{12}+\zeta_{12}&-i\omega m_{22}+\zeta_{22}
\end{array}\right),
\end{equation}
variable vector
\begin{equation}
\label{3.4}\vc{\psi}_\omega=(q_{1\omega},q_{2\omega},U_{1\omega},U_{2\omega}),
\end{equation}
and force vector
\begin{equation}
\label{3.5}\vc{\varphi}_\omega=(0,0,E_{1\omega},-E_{1\omega}).
\end{equation}
The solution of Eq. (3.2) is given by
\begin{equation}
\label{3.6}\vc{\psi}_\omega=\vc{G}(\omega)\cdot\vc{\varphi}_\omega,\qquad\vc{G}(\omega)=\vc{M}(\omega)^{-1}.
\end{equation}
It is convenient to introduce the $2\times 2$ impedance matrix
\begin{equation}
\label{3.7} \du{z}(\omega)=\vc{\zeta}-i\omega \du{m}.                                                                                                                                    \end{equation}

From the solution vector $\vc{\psi}_\omega$ we can construct the corresponding displacements
\begin{equation}
\label{3.8}d_{j\omega}=q_{j\omega}+\frac{U_{\omega}}{i\omega},\qquad (j=1,2),
\end{equation}
where $U_\omega=(U_{1\omega}+U_{2\omega})/2$. It follows that
\begin{equation}
\label{3.9}2i\omega d_{1\omega}=U_{{2\omega}}-U_{{1\omega}},\qquad  2i\omega d_{2\omega}=U_{{1\omega}}-U_{{2\omega}}.
\end{equation}

Substituting Eq. (2.10) into (2.8) we see that the mechanical mean swimming velocity is linear in the four correlations $\overline{d^{(1)}_jU^{(1)}_k}$.
From the solution Eq. (3.6) we find the equalities
\begin{equation}
\label{3.10}\overline{d^{(1)}_1U^{(1)}_1}=\overline{d^{(1)}_1U^{(1)}_2}=-\overline{d^{(1)}_2U^{(1)}_1}=
-\overline{d^{(1)}_2U^{(1)}_2}.
\end{equation}
Using this we obtain
\begin{equation}
\label{3.11}\overline{U^{(2)}}_M=2\frac{Z_{0d}}{Z_0} \;\overline{d^{(1)}_1U^{(1)}_1},
\end{equation}
with derivative with respect to $d$,
\begin{equation}
\label{3.12}Z_{0d}=\partial Z_0/\partial d,
\end{equation}
taken at $z=d$. In complex notation we use that for two real quantities $A(t)$ and $B(t)$ oscillating at frequency $\omega$,
\begin{equation}
\label{3.13}A(t)=\mathrm{Re}[A_\omega e^{-i\omega t}],\qquad B(t)=\mathrm{Re}[B_\omega e^{-i\omega t}],
\end{equation}
the time-average of the product is given by
\begin{equation}
\label{3.14}\overline{A(t)B(t)}=\frac{1}{2}\mathrm{Re}[A_\omega^*B_\omega].
\end{equation}

For the Oseen* and Oseen-Dipole models Eq. (3.11) agrees with the expression for the mean swimming velocity derived by a different method \cite{4}. The mean swimming velocity must be supplemented with the additional contribution from the motion of the fluid.. First we show how the above derivation must be modified when the friction coefficients depend on frequency, and find a general expression for the moment in Eq. (3.11).

\section{\label{IV}Frequency-dependent friction}

It was first shown by Stokes \cite{5} on the basis of the linearized Navier-Stokes equations that the flow pattern about an oscillating sphere differs from that about a steadily moving sphere. Correspondingly the friction coefficient of a single sphere depends on frequency. For a sphere of radius $a$ with no-slip boundary condition it takes the value $\zeta_0=6\pi\eta a$ at zero frequency, but at higher frequency the complex coefficient $\zeta(\omega)$ differs. It was shown by Basset \cite{6} that the frequency-dependent friction corresponds to a retardation effect in the hydrodynamic force exerted by the fluid. If the fluid is viscoelastic, then that is an additional reason for retardation. In the following we discuss how the retardation effect can be taken into account in the theory of swimming.

We modify the equations of motion Eq. (2.1) to include retarded friction, and replace the frictional term $-\vc{\zeta}\cdot\du{U}$ by Basset-type forces $\du{B}(t)$ with
 \begin{equation}
\label{4.1}B_i(t)=-\int^t_{-\infty} \beta_{ij}(z_2(t')-z_1(t'),t-t')U_j(t')\;dt',\;\qquad (i,j)=(1,2)
\end{equation}
with memory kernel $\beta_{ij}(z,\tau)$. The momentum balance Eq. (2.3) is modified correspondingly to
 \begin{equation}
\label{4.2}\frac{dP}{dt}=-\du{u}\cdot\int^\infty_0\vc{\beta}(\du{z}(t-\tau),\tau)\cdot\du{U}(t-\tau)\;d\tau.
\end{equation}
A $\delta'(t-\tau)$-contribution to the kernel takes account of the added mass effect \cite{7}.

We consider the asymptotic periodic swimming motion. Integrating Eq. (4.2) over a period $T=2\pi/\omega$ we obtain

 \begin{equation}\label{4.3}\int^T_0dt\;\du{u}\cdot\int^\infty_0\vc{\beta}(\du{z}(t-\tau),\tau)\cdot\du{U}(t-\tau)\;d\tau=0.
\end{equation}
To lowest order the swimmer is at rest with positions $\du{z}_0$ and vanishing velocities $\du{U}^{(0)}=0$. To first order the motion is oscillatory and Eq. (4.3) is satisfied because
 \begin{equation}
\label{4.4}\int^T_0dt\;\du{U}^{(1)}(t-\tau)=0.
\end{equation}
To second order Eq. (4.3) yields an equation for the mechanical mean swimming velocity, in generalization of Eq. (2.8). The equation reads
\begin{equation}
\label{4.5}\overline{U^{(2)}}_M=\frac{-1}{Z_0}\;\int^T_0dt\;\du{u}\cdot\int^\infty_0\vc{\beta}^{(1)}(\du{z}_0,\tau)\cdot\du{U}^{(1)}(t-\tau)\;d\tau.
\end{equation}
This shows that the mechanical mean second order swimming velocity can be calculated from the linear response. The equations are the same as before, except that the friction matrix has become frequency-dependent.

In the linear theory the equations take the form  (3.2) with frequency-dependent complex friction coefficients $\zeta_{ij}(\omega)$, and masses $m_{ij}$. The added masses are defined by
\begin{equation}
\label{4.6}m_{ij}=m_i\delta_{ij}+m_{aij},
\end{equation}
 where $m_1,m_2$ are the bare masses. These definitions imply that the friction coefficients do not have an imaginary part proportional to $\omega$ at high frequency. Since only the impedance matrix $\du{z}(\omega)$ is relevant, we could alternatively use the bare masses only, and include the added masses in the friction coefficients.

The property
 \begin{equation}
\label{4.7}z_{ij}(-\omega)=z_{ij}(\omega)^*,\qquad (i,j)=(1,2)
\end{equation}
leads to the same mathematical structure as in Sec. III with a $4\times 4$ matrix $\vc{M}(\omega)$ with the property (3.7). As a consequence, for real activating forces $E_1(t)=E_{1c}\cos(\omega t),\;E_2(t)=-E_1(t),$ the solution $\vc{\psi}(t)$ takes the form
\begin{equation}
\label{4.8} \vc{\psi}(t)=\vc{\psi}_{\omega c}\cos(\omega t)+\vc{\psi}_{\omega s}\sin(\omega t)
\end{equation}
with real vectors $\vc{\psi}_{\omega c}=\mathrm{Re}[\vc{\psi}_\omega]$ and $\vc{\psi}_{\omega s}=\mathrm{Im}[\vc{\psi}_\omega]$ which follow from the solution Eq. (3.6) with the new matrix $\vc{M}(\omega)$ and with $\vc{\varphi}_\omega$ determined from $E_1,E_2=-E_1$. We put $E_{1c}=\varphi ka$, where $\varphi$ is a dimensionless amplitude.

The complex velocity amplitudes are found from the linear equations as
\begin{eqnarray}
\label{4.9}U_{1\omega}&=&\frac{-i \omega\varphi k a}{\Delta(k,\omega)}(z_{12}+z_{22}),\nonumber\\
U_{2\omega}&=&\frac{-i\omega\varphi k a}{\Delta(k,\omega)}(-z_{12}-z_{11}),
\end{eqnarray}
with denominator
\begin{equation}
\label{4.10}\Delta(k,\omega)=k Z(\omega)-i\omega|\du{z}|,
\end{equation}
where $|\du{z}|$ is the determinant of the complex impedance matrix $\du{z}$. The corresponding displacement amplitudes are
\begin{equation}
\label{4.11}d_{1\omega}=-d_{2\omega}=\frac{\varphi kaZ(\omega)}{2\Delta(k,\omega)}.
\end{equation}

The expression for the moment in Eq. (3.11) becomes
\begin{equation}
\label{4.9}\overline{d^{(1)}_1U^{(1)}_1}=\frac{1}{4}\varphi^2k^2a^2\omega\;\frac{N(\omega)}{D(k,\omega)},
\end{equation}
with numerator
\begin{equation}
\label{4.10}N(\omega)=-z''_{11}(z'_{12}+z'_{22})+z''_{12}(z'_{11}-z'_{22})+z''_{22}(z'_{12}+z'_{11}),
\end{equation}
and denominator
\begin{equation}
\label{4.11}D(k,\omega)=|\Delta(k,\omega)|^2.
\end{equation}

The numerator $N(\omega)$ can be expressed as
\begin{equation}
\label{4.15} N(\omega)=\frac{1}{2}\du{u}\cdot\du{z}^*(\omega) \cdot\vc{\sigma}_y\cdot\du{z}(\omega)\cdot\du{u},
\end{equation}
where $\vc{\sigma}_y$ is the Pauli spin-matrix \cite{12},
\begin{equation}
\label{4.16}\vc{\sigma}_y=\left(\begin{array}{cc}0&-i
\\i&0
\end{array}\right).
\end{equation}
The sign of $N(\omega)$ determines the direction of swimming.
The absolute value $|N(\omega)|$ equals twice the area of the triangle in the $z'z''-$ plane made up by the points $P_0=(-z'_{12},-z''_{12})$, $P_1=(z'_{11},z''_{11})$, $P_2=(z'_{22},z''_{22})$, as given by the so-called Shoelace Formula \cite{13}$^,$\cite{14}. We call the triangle the impedance triangle. With $3-$vectors $\vc{P}_0=(P_0,0),\;\vc{P}_1=(P_1,0),\;\vc{P}_2=(P_2,0)$, and difference vectors $\vc{S}_1=\vc{P}_1-\vc{P}_0$  and $\vc{S}_2=\vc{P}_2-\vc{P}_0$, the value $N(\omega)$ is the third component of the cross product $\vc{N}(\omega)=\vc{S}_1\times\vc{S}_2$. The vector $\vc{N}(\omega)$ can be regarded as a generalized torque corresponding to the impedance matrix $\du{z}(\omega)$.

We wish to choose the amplitude factor $\varphi$ in such a way that the mean square displacement does not vary with the scale number $s$. It follows from Eq. (3.9) that $d^{(1) }_2=-d^{(1)}_1$, so that it suffices to consider $ \overline{(d_1^{(1)})^2}$.
 For $E_1(t)=\varphi ka\cos(\omega t)$ the mean square displacement is calculated as
 \begin{equation}
\label{4.17}\overline{(d_1^{(1)})^2}=\frac{1}{8}\varphi^2k^2a^2\;\frac{|Z(\omega)|^2}{D(k,\omega)}.
\end{equation}
If we choose
\begin{equation}
\label{4.18}\varphi ^2=\varepsilon^2\frac{D(k,\omega)}{k^2|Z(\omega)|^2},
\end{equation}
then $\overline{(d_1^{(1)})^2}$ takes the constant value $\varepsilon^2a^2/8$, and the moment in Eq. (4.12) becomes
\begin{equation}
\label{4.19}\overline{d^{(1)}_1U^{(1)}_1}=\frac{1}{4}\varepsilon^2a^2\omega\;\frac{N(\omega)}{|Z(\omega)|^2}.
\end{equation}
In combination with Eq. (3.11) we therefore find for the mechanical contribution to the mean swimming velocity
\begin{equation}
\label{4.20}\overline{U^{(2)}}_M=\varepsilon^2a^2\omega\frac{Z_{0d}}{2Z_0} \;\frac{N(\omega)}{|Z(\omega)|^2}.
\end{equation}
We note that this is independent of the elasticity coefficient $k$. The dependence on $k$ has been fully absorbed in the expression (4.18) for the amplitude factor $\varphi$.

In the same way we calculate the averages $\overline{\big(U^{(1)}_1\big)^2}$, $\overline{\big(U^{(1)}_2\big)^2}$, and  $\overline{U^{(1)}_1U^{(1)}_2}$. We find
\begin{equation}
\label{4.21}\overline{\big(U^{(1)}_j\big)^2}=\frac{1}{2}
\varepsilon^2 a^2\omega^2\;\frac{N_ j(\omega)}{|Z(\omega)|^2},\qquad \overline{U^{(1)}_1U^{(1)}_2}=-\frac{1}{2}\varepsilon^2a^2\omega^2\frac{N_ 3(\omega)}{|Z(\omega)|^2},
\end{equation}
with numerators
\begin{eqnarray}
\label{4.22}N_1(\omega)&=&
(z'_{12}+z'_{22})^2+(z''_{12}+z''_{22})^2,\nonumber\\
N_2(\omega)&=&
(z'_{12}+z'_{11})^2+(z''_{12}+z''_{11})^2,\nonumber\\
N_3(\omega)&=&-
(z'_{12}+z'_{11})(z'_{12}+z'_{22})+(z''_{12}+z''_{11})(z''_{12}+z''_{22}).
\end{eqnarray}
Again the averages are independent of the elasticity coefficient.

\section{\label{V}Effectiveness}

In this section we study the frequency-dependence of the mean swimming velocity in more detail. We use the dimensionless number $R=a^2\omega\rho/\eta=2s^2$ as a measure of frequency. We have $R=\omega\tau_v$ where $\tau_v=a^2\rho/\eta$ is the viscous relaxation time. In the notation of Derr et al.\cite{16} $R=M^2$. We define the dimensionless effectiveness $V(R)$ as
\begin{equation}
\label{5.1}\overline{U_{sw}}=\varepsilon^2 a\omega V(R).
\end{equation}
In words, $\varepsilon^2V(R)=\overline{U_{sw}}/(a\omega)$ is the ratio of the distance traveled during a period $T=2\pi/\omega$ to the length $2\pi a$, when swimming with a stroke of amplitude $\varepsilon$ at scale number $R$. Clearly one wishes to optimize this value.
We recall that the efficiency, as defined by Shapere and Wilczek\cite{16} involves both the mean swimming velocity and the power dissipated during a period. The effectiveness of the stroke is defined regardless of the power.

The mechanical contribution to the effectöiveness $V_M(R)$ in the limit $\varepsilon\rightarrow 0$ is found by combination of Eqs. (4.20) and (5.1) as
\begin{equation}
\label{5.2}V_M(R)=\frac{aZ_{0d}}{2Z_0} \;\frac{N(\omega)}{|Z(\omega)|^2}.
\end{equation}
This takes a simple form for models with frequency-independent friction matrix. In that case we find from Eq. (4.15)
\begin{equation}
\label{5.3}N(\omega)=N'\omega,
\end{equation}
with coefficient
\begin{equation}
\label{5.4}N'=\frac{i}{2}\;\du{u}\cdot(\du{m}\cdot\vc{\sigma}_y\cdot\vc{\zeta}-\vc{\zeta}\cdot\vc{\sigma}_y\cdot\du{m})\cdot\du{u},
\end{equation}
given by
\begin{eqnarray}
\label{5.5}N'&=& m_{11}(\zeta_{12}+\zeta_{22})-m_{12}(\zeta_{11}-\zeta_{22})-m_{22}(\zeta_{12}+\zeta_{11})\nonumber\\
&=&\frac{1}{2}(Z_0\Delta m-M_0\Delta\zeta),
\end{eqnarray}
with differences
\begin{equation}
\label{5.6} \Delta m=m_{11}-m_{22},\qquad \Delta\zeta=\zeta_{11}-\zeta_{22}.
\end{equation}
This shows an interesting symmetry between mass and friction.

We find that $V_M(R)$ takes the form \cite{4}
\begin{equation}
\label{5.7}V_{M2}(R)=V_{xM}\frac{2R_{xM}R}{R_{xM}^2+R^2},
\end{equation}
with parameters $R_{xM},V_{xM}$. The function $V_{M2}(R)$ has maximum value $V_{xM}$ at $R_{xM}$, and the curve has half-width $\Delta R_=2\sqrt{3}R_{xM}$. We call a curve in the $VR$-plane with behavior described by Eq. (5.7) a $VR2$-curve.

From the behavior at small and large $R$
\begin{eqnarray}
\label{5.8)}V_{M2}(R)&\approx& CR \qquad\mathrm{as}\;R\rightarrow 0,\nonumber\\ V_{M2}(R)&\approx& L/R \qquad\mathrm{as}\;R\rightarrow \infty,
\end{eqnarray}
we find
\begin{equation}
\label{5.9}R_{xM}=\sqrt{L/C},\qquad V_{xM}=\frac{1}{2}\sqrt{LC}.
\end{equation}
From Eq. (4.20) we find the values
\begin{equation}
\label{5.10}C=\frac{aZ_{0d}}{2Z_0}\;\frac{\eta |N'|}{a^2\rho Z_0^2},\qquad L=\frac{aZ_{0d}}{2Z_0}\;\frac{a^2\rho |N'|}{\eta M^2},
\end{equation}
where $M=\du{u}\cdot\du{m}\cdot\du{u}=m_{11}+2m_{12}+m_{22}$ is the total mass.
Therefore
\begin{equation}
\label{5.11}R_{xM}=\frac{a^2\rho Z_0}{\eta M},\qquad V_{xM}=\frac{aZ_{0d}}{4Z_0}\frac{|N'|}{MZ_0}.
\end{equation}

The above expressions can be cast in the form \cite{4}
\begin{equation}
\label{5.12}V_{M2}(R)=\mathrm{Im}\frac{\chi_0}{1-i\omega\tau_M},
\end{equation}
with
\begin{equation}
\label{5.13}\tau_M=\frac{\tau_v}{R_{xM}}=\frac{M}{Z_0},\qquad \chi_0=\frac{2V_{xM}}{R_{xM}}.
\end{equation}
This shows that for such models the dependence on frequency is quite simple.

In earlier work \cite{4} we derived explicit expressions for the parameters $R_{xM},V_{xM}$ for three specific models. The Oseen-model is based on the Oseen approximation for the hydrodynamic pair mobility and added mass effects are neglected. The Oseen$^*$-model uses the same approximation for the pair mobility, and includes the single sphere added mass effect. The Oseen-Dipole model describes in addition the pair added mass in dipole approximation.  We note that for the Oseen$^*$-model $M=m_1^*+m_2^*$, where $m^*_j=m_j+\frac{1}{2}m_{fj}$ with $m_{fj}=(4\pi/3)\rho a_j^3$. For the Oseen-Dipole model $M=m_{11}+2m_{12}+m_{22}$ with $m_{ij}$ given below. The factors $Z_{0d}$ and $Z_0$ are the same for the three models. The Oseen mobility matrix corresponds to the
friction matrix
\begin{equation}
\label{5.14}\vc{\zeta}=\frac{12\pi\eta z}{4z^2-9ab}\left(\begin{array}{cc}2az&-3ab\\-3ab&
2bz\end{array}\right),
\end{equation}
where $z=z_2-z_1$. In the Oseen-Dipole model one uses a dipole approximation to the mass matrix given by \cite{11}
\begin{equation}
\label{5.15}\du{m}=\frac{2\pi}{3z^6-3a^3b^3}\left(\begin{array}{cc}a^3[z^6(\rho+2\rho_a)
+2a^3b^3(\rho-\rho_a)]&-3a^3b^3z^3\rho\\-3a^3b^3z^3\rho&
b^3[z^6(\rho+2\rho_b)+2a^3b^3(\rho-\rho_b)]\end{array}\right).
\end{equation}
 In an expansion in inverse powers of $z$ to terms of order $z^{-6}$ this agrees with the expression given by Lamb \cite{21}. The diagonal elements include the single sphere added mass effect.

The mechanical contribution to the effectiveness, given by Eq. (5.2), can be evaluated straightforwardly for the three models  by use of the above expressions for $\vc{\zeta}$ and $\du{m}$. We find agreement with those found before \cite{4}, except for a copying error in the coefficient $r_b$ in Eq. (6.16) of Ref. 4. This should read instead
\begin{equation}
\label{5.17}r_b=2\xi^2(\xi-\delta^2)(2\delta-3\xi)(\delta^4+\delta^2\xi+\xi^2),
\end{equation}
where $\xi=b/a$ and $\delta=d/a$.

In particular we find for the Oseen-model
\begin{equation}
\label{5.18}V_{MO}(R)=\frac{27}{4}s^2\;\frac{(2\delta-3)(2\delta-3\xi)}{\delta-3´\xi+\delta\xi}\;\frac{(2\delta-3)\sigma_a-(2\delta-3\xi)\xi^2\sigma_b}{81\delta^2(\delta-3\xi+\delta\xi)^2
+(4\delta^2-9\xi)^2(\sigma_a+\xi^3\sigma_b)^2s^4},
\end{equation}\
where $\sigma_a=\rho_a/\rho$ and  $\sigma_b=\rho_b/\rho$ .
We compare with the expression for the effectiveness in the Oseen-model, as calculated by Hubert et al. \cite{3} for large $d/a$. Their Eq. (SM 18) reads in our notation
\begin{equation}
\label{5.19}\overline{V_M}^S=\frac{3\xi^2}{2(1+\xi)^3\delta^2}\;\frac{\omega(\tau_a-\tau_b)}{1+\omega^2\tau_{MO}^2},
\end{equation}
where $\tau_a=m_1/(6\pi\eta a),\; \tau_b=m_2/(6\pi\eta b),\;\tau_{MO}=(m_1+m_2)/(6\pi\eta (a+b))$. This agrees to order $\delta^{-2}$ with Eq. (5.17). For the Oseen$^*$-model the same equations hold with $m_i$ replaced by $m_i^*=m_i+(2\pi/3)\rho a_i^3$, or with $\sigma_i$ replaced by $\sigma_i^*=\sigma_i+\frac{1}{2}$.

The numerator of the last factor in Eq. (5.17) for fixed $\sigma_a,\xi,\delta$ can switch sign as a function of $\sigma_b$. For example for $\sigma_a=1,\;\xi=1/2,\;\delta=3$ the expression switches sign at $\sigma_b=8/3$. This corresponds to a change of swimming direction.

For the Oseen-model the maximum value $V_{xMO}$ depends on the mass densities $\rho_a,\rho_b$ only via the ratio $r=\rho_b/\rho_a=\sigma_b/\sigma_a$. The expression is, from Eq. (5.11),
\begin{equation}\label{5.20}V_{xMO}=\frac{3\xi^2(2\delta-3)(2\delta-3\xi)(3-2\delta+2r\delta\xi^2-3r\xi^3)}{8\delta(4\delta-9\xi^2)(\delta-3\xi+\delta\xi)^2(1+q\xi^3)}.
\end{equation}
To get the value for the Oseen$^*$-model replace $r$ by $r^*=\sigma_b^*/\sigma_a^*$. For the position of the maximum one finds
\begin{equation}\label{5.21}R_{xMO}=\frac{18\delta(\delta-3\xi+\delta\xi)}{(4\delta^2-9\xi)(1+r\xi^3)}\;\frac{1}{\sigma_a},
\end{equation}
and similarly for $R_{xMS}$. These simple properties do not hold for the Oseen-Dipole model. For typical values of $\xi,\delta$ the value of $V_{xMO}$ ranges from positive to negative values. From Eq. (5.19) one sees that in the Oseen-model the sign switch occurs at
\begin{equation}\label{5.22}r_0=\frac{2\delta-3}{\xi^2(2\delta-3\xi)}.
\end{equation}\\
 As a standard system we choose a two-sphere with parameters $b=\frac{1}{2}a,\;d=3a$ in a fluid with viscosity $\eta$ and mass density $\rho$. The mass densities of the spheres can be varied, and take values $\rho_a=\sigma_a\rho$ and $\rho_b=\sigma_b\rho$.
We consider first neutrally buoyant spheres with $\sigma_a=1,\;\sigma_b=1$. In Fig. 1 we show the effectiveness $V_{M}$ as a function of $R$ for the Oseen-model, the Oseen$^*$- model, and the Oseen-Dipole model. In the notation introduced at the end of Sec. I the three models are indicated respectively as the $NO$-model, the $NS$-model, and the $ND$-model. In  Table I we list the values of the parameters $R_{xM}$ and $V_{xM}$ for the three models.\\

\setlength{\unitlength}{1cm}
\begin{figure}
 \includegraphics{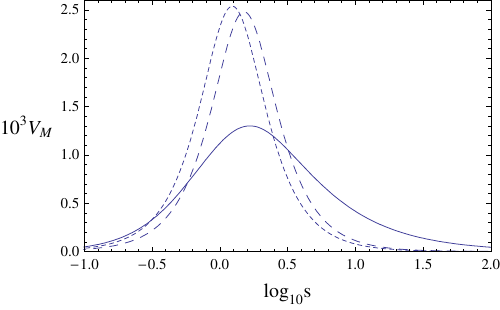}
   \put(-9.1,3.1){}
\put(-1.2,-.2){}
  \caption{Plot of the mechanical effectiveness $V_M$ for a two-sphere with $a=1$, $b=1/2$, $d=3$, $\rho_a=\rho_b=\rho$, with hydrodynamic interactions of Oseen-tye (NO: long dashes), , dipole-type added mass (ND: short dashes). retarded friction (NR: solid curve). The curve for NS is not shown, since it cannot be distinguished from the one for ND}
\end{figure}

\begin{table}[!htb] \footnotesize\centering

  \caption{List of values characterizing the mechanical swimming speed of a two-sphere with radii $a=1,\;b=1/2$, equilibrium distance between centers $d=3$, in a fluid with viscosity $\eta=1$, mass density $\rho=1$, for four models: Oseen (O), Oseen*(S), dipole (D), retarded friction (R), of two systems $N=(\rho,\rho)$ or $L=(\rho,16\rho)$. The listed values are explained in Sec. V. In the case of retarded friction the relations in Eq. (5.11) do not hold.}\label{tab:1}

\begin{tabular}{|c|c|c|c|c|c|}
\hline
 $(\rho_a,\rho_b)$ &$model$ &$R_{2xM}$ &$10^6V_{2xM}$ &$Z_0/(\eta a)$ &$M/(\rho a^3)$
 \rule[-5pt]{0pt}{16pt} \\
\hline
 $(\rho,\rho)$ &$Oseen$ &$4.571$ &$2480$ &$21.5423$ &$4.712$
 \rule[-5pt]{0pt}{16pt} \\
 \hline
  $(\rho,\rho)$ &$Oseen^*$ &$3.048$ &$2480$ &$21.5423$ &$7.069$
  \rule[-5pt]{0pt}{16pt} \\
  \hline  $(\rho,\rho)$ &$Dipole$ &$3.072$ &$2538$ &$21.5423$ &$7.012$
  \rule[-5pt]{0pt}{16pt} \\
  \hline  $(\rho,\rho)$ &$RetFr$ &$5.255$ &$1298$ &$21.7598$ &$7.014$
  \rule[-5pt]{0pt}{16pt} \\
  \hline  $(\rho,16\rho)$ &$Oseen$ &$1.714$ &$-7440$ &$21.5423$ &$12.566$
  \rule[-5pt]{0pt}{16pt} \\
  \hline  $(\rho,16\rho)$ &$Oseen^*$ &$1.444$ &$-5874$ &$21.5423$ &$14.923$
  \rule[-5pt]{0pt}{16pt} \\
  \hline
    $(\rho,16\rho)$ &$Dipole$ &$1.449$ &$-5879$ &$21.5423$ &$14.866$
  \rule[-5pt]{0pt}{16pt} \\
  \hline
    $(\rho,16\rho)$ &$RetFr$ &$4.409$ &$-2827$ &$21.7598$ &$14.868$
  \rule[-5pt]{0pt}{16pt} \\
  \hline
\end{tabular}

\end{table}

For our choice of parameters $\xi=1/2,\;\delta=3$, we have $R_{xMNO}=32/7=4.571$ and $V_{xMNO}=5/2016=0.002480$. As a function of $\log_{10}R$ the curve resembles a gaussian. For the $NS$-model the corresponding numbers read: $R_{xMNS}=64/21=3.048$ and $V_{xMNS}=5/2016=0.002480$.
 We compare with the curve $V_{MND}(R)$ for the $ND$-model, characterized by the same friction matrix Eq. (5.14), but with mass matrix given by Eq. (5.15). The function $V_{MND}(R)$ has the form Eq. (5.7) with parameter values $R_{xMND}=3.063$ and $V_{xMND}=0.002538$. Thus the two $VR2-$curves are nearly identical.

  For the $NS$- model the points $P_j$  of the impedance triangle, defined below Eq. (4.16), have coordinates $P_0=(5.386 \eta a,0)$, $P_1=( 21.542\eta a,-19.149\omega\rho a^3)$, $P_2=( 10.771\eta a, -2.394\omega\rho a^3)$ at $R=R_{xMO}$. The value of the coefficient $N'$ is $N'=21.15\eta\rho a^4$. For the $ND$-model the points are $P_0=(5.386\eta a,-0.089\omega\rho a^3), \;P_1=(21.542\eta a,-19.308\omega\rho a^3), \;P_2=(10.771\eta a,-2.413\omega\rho a^3)$, and $N'=21.47\eta\rho a^4$.

 As a second example we consider the $L$-system with
 the same geometry, but with mass densities $\rho_a=\rho$ and $\rho_b=16\rho$, so that $m_2=2m_1$. For this geometry $r_0=8/3$, so that the $r$-values of the two systems, $r=1$ and $r=16$, lie on opposite sides of $r_0$. Hence the two systems swim in opposite directions. We list the corresponding values in Table I. Both systems were studied also by Derr et al.\cite{15}.

\section{\label{VI}Retarded friction and added mass}

In this section we study the effect of frequency-dependent friction.
We do not aim at a complete calculation based on exact frequency-dependent friction coefficients, but limit ourselves to the dominant effects for not too small distance between centers. At very large distance the interaction tends to zero and the friction matrix is diagonal with single-sphere terms given by the frequency-dependent friction coefficient first calculated by Stokes \cite{5}.

The mutual mobility will be approximated by a result\cite{8}, based on a single Green function between the spheres and on Fax\'en theorems. We have used the approximation earlier in a study of velocity relaxation of a pair of spheres \cite{17}. In the limit of zero frequency the approximation reduces to that of Rotne and Prager \cite{18}, as generalized to spheres of different sizes \cite{19}. The Rotne-Prager approximation is popular in soft matter modeling \cite{20}. The self-mobility is given by the single-sphere friction coefficient, corrected by a distance-dependent term, as detailed below, chosen such that the mass matrix is identical with that of the Oseen-Dipole model, given by Eq. (5.15).

The single sphere friction coefficient of spheres $A$ and $B$ is, respectively,
\begin{equation}
\label{6.1}\zeta_A(\omega)=6\pi\eta aA_0(\alpha a),\qquad\zeta_B(\omega)=6\pi\eta bA_0(\alpha b),
\end{equation}
with the abbreviation \cite{7}
\begin{equation}
\label{6.2}A_0(\lambda)=1+\lambda+\frac{1}{9}\lambda^2,
\end{equation}
and $\alpha=\sqrt{-i\omega\rho/\eta},\;\mathrm{Re}[\alpha]>0$.
The approximate mutual mobility function for two spheres with distance $z$ between centers reads \cite{8}
\begin{equation}
\label{6.3}\alpha_{AB}^{tt}(z,\omega)=\frac{B_0(\alpha a) B_0(\alpha b)-(1+\alpha z)
e^{\alpha(a+b-z)}}{2\pi\eta\alpha^2A_0(\alpha a)A_0(\alpha b)z^3},
\end{equation}
with
\begin{equation}
\label{6.4}
B_0(\lambda)=1+\lambda+\frac{1}{3}\lambda^2.
\end{equation}

The pair mobility matrix will be approximated by
\begin{equation}
\label{6.5} \vc{\mu}(z,\omega)=\left(
\begin{array}{cc}\alpha^{tt}_{AA}(z,\omega)&\alpha^{tt}_{AB}(z,\omega)
\\\alpha^{tt}_{AB}(z,\omega)&\alpha^{tt}_{BB}(z,\omega)
\end{array}\right),
\end{equation}
with the off-diagonal terms given by Eq. (6.3). The diagonal terms are given by the single-sphere terms with a Van der Waals-type pair correction, corresponding to a dipolar reflection of a dipole field,
\begin{equation}
\label{6.6}
\alpha^{tt}_{AA}(z,\omega)=
\frac{1}{6\pi\eta aA_0(z,\omega)},\qquad A_0(z,\omega)=1+\alpha a+\frac{1}{9}\bigg(1-6\frac{a^3b^3}{z^6}\bigg)\alpha^2a^2.
\end{equation}
The approximate pair friction matrix is
\begin{equation}
\label{6.7}\vc{\zeta}(z,\omega)=\vc{\mu}(z,\omega)^{-1}.
\end{equation}
.The real and imaginary parts of the matrix elements are denoted as $\zeta'_{ij}´$ and $\zeta''_{ij}$.

The pair friction matrix given by Eqs. (6.1-7) increases with frequency in proportion to $-i\omega$. It therefore contains added mass terms, as explained below Eq. (4.6). We incorporate all added mass effects in the friction matrix and use the impedance matrix
\begin{equation}
\label{6.8} \du{z}(\omega)=\vc{\zeta}-i\omega\du{m}_0,                                                                                                                                      \end{equation}
rather than Eq. (3.7). With $\vc{\zeta}(z,\omega)$ given by Eqs. (6.1-7) the added mass matrix is identical with that given by Eq. (5.15).

In the expression (5.2) for the function $V_{M}(R)$ we need the pre-factor $aZ_{0d}/Z_0$, where $Z_0=\du{u}\cdot\vc{\zeta}(0)\cdot\du{u}$ is the steady-state friction coefficient, and $Z_{0d}$ its derivative with respect to $d$. With the above expressions we find
\begin{equation}
\label{6.9}Z_0=24\pi\eta d^3\frac{(a+b) d^3-3 a b d^2+a b (a^2+b^2)}{4 d^6-9 a b d^4+6 a b (a^2+b^2) d^2-a b (a^4+b^4)-2 a^3 b^3}.
\end{equation}
This is identical with the Rotne-Prager-approximation.
For large $d$,
\begin{eqnarray}
\label{6.10}\frac{Z_{0d}}{Z_0}&=&\frac{3 a b}{(a+b) d^2}-\frac{9 a b(a^2+b^2)}{ 2(a+b)^2d^3}\nonumber\\
&+&\frac{3ab}{(a+b)^3d^4}(a^4+2a^3b-7a^2b^2+2ab^3+b^4)+O(d^{-5}).
\end{eqnarray}
To the order shown this is identical with the exact result \cite{19}.

In Fig. 2 we show the mechanical effectiveness $V_{M}(R)$ for the $RF$-model as specified above. For the $N$-system the maximum is at $R_{mM                                                NR}=5.573$ and has value $V_{mMNR}=0.001300$. We compare with the $VR2$-curve for these parameter values. It is evident that the retardation of friction has a significant effect. In a plot of $V$ vs. $R$ the half-width of the actual curve is $ \Delta R=73.28$, whereas for the $VR2-$curve $ \Delta R_2=2\sqrt{3}R_{mMNR}=19.30$. The values for the minimum for the $L$-system with $m_2=2m_1$ are $R_{mMLR}=4.389$ and $V_{mMLR}=-0.002828$.  The half-width of the spectral curve is $ \Delta R=64.02$, whereas for the $VR2-$curve $ \Delta R_2=2\sqrt{3}R_{mMLR}=15.20$. Hence in both cases the retarded friction causes a shift to higher frequency of the extremum and a strong broadening of the spectral curve. In Fig. 3 we show the analogue of Fig. 1 for the $L$-system, and in Fig. 4 we show the analogue of Fig. 2.

\newpage
\setlength{\unitlength}{1cm}
\begin{figure}
 \includegraphics{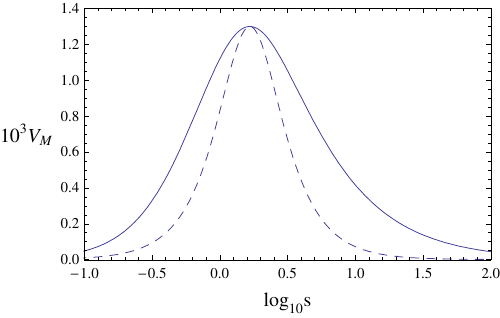}
   \put(-9.1,3.1){}
\put(-1.2,-.2){}
  \caption{Plot of mechanical effectiveness $V_M$ for the $NR$-model (solid curve) compared with curve of the form (5.7) for the same maximum (dashed).\\}
\end{figure}

\setlength{\unitlength}{1cm}
\begin{figure}
 \includegraphics{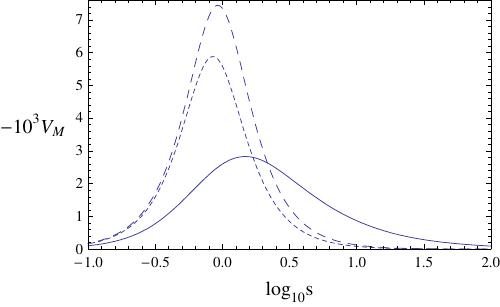}
   \put(-9.1,3.1){}
\put(-1.2,-.2){}
  \caption{Plot of the mechanical effectiveness $V_M$ for  a two-sphere with $a=1$, $b=1/2$, $d=3$, $\rho_a=\rho,\;\rho_b=16\rho$, with hydrodynamic interactions of Oseen-tye (LO: long dashes) dipole-type added mass (LD: short dashes), retarded friction (LR: solid curve). The curve for LS is not shown, since it cannot be distinguished from the one for LD.
}
\end{figure}
\newpage
\setlength{\unitlength}{1cm}
\begin{figure}
 \includegraphics{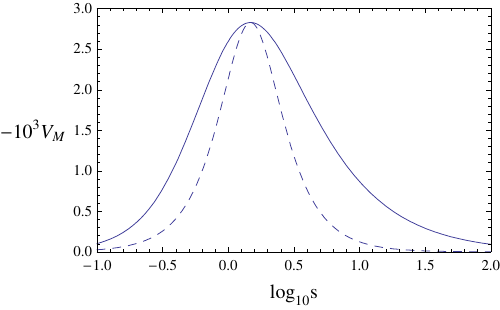}
   \put(-9.1,3.1){}
\put(-1.2,-.2){}
  \caption{Plot of the mechanical effectiveness $V_M$ for the $LR$-model (solid curve) compared with curve of the form (5.7) for the same minimum (dashed).}
\end{figure}

\clearpage
\newpage

\section{\label{VII}Discussion}

In the above we analyzed the mechanical contribution to the mean swimming velocity of a two-sphere swimmer. We showed that this contribution can be expressed in terms of the impedance matrix of the two-sphere. The latter embodies the linear response of the two--sphere to applied oscillatory forces acting on either sphere. The response depends strongly on frequency. We studied this dependence for various, more or less realistic choices of the hydrodynamic interactions incorporating the effects of friction and added mass.

Part of the desired information on the mean swimming velocity is missing. We must yet add the hydrodynamic contribution due to fluid flow generated by the oscillatory motion of the two spheres and affected by the Reynolds stress in the fluid. Many aspects of the required calculations were discussed by Derr et al. \cite{15}. In separate work we shall compare the hydrodynamic and mechanical contributions. Once a reliable calculation of the mean swimming velocity of the vibrating two-sphere is established we can apply the method to other systems of interest, for example to a vibrating chain of three spheres.\\\\

The author has no conflicts to disclose.

\newpage
\end{document}